\documentclass[useAMS,usenatbib,useverbatim]{mn2e}
\usepackage{graphicx}
\usepackage{psfig}
\bibpunct{(}{)}{;}{a}{}{,}    
\usepackage{amsmath}
\usepackage{amssymb,graphicx}
\usepackage{verbatim}

\def\me{$\dot{m}_{E}$}

\usepackage{journal_names}

\title[Radio/X-ray variability in NGC~4051]{ Radio and X-ray
  variability in the Seyfert galaxy NGC~4051} 

\author[S.Jones et al]{S.Jones$^{1}$\thanks{E-mail:
s.jones@phys.soton.ac.uk (SJ)}, I.M$^{\mbox{c}}$Hardy$^{1}$, D.Moss$^{1}$,
N.Seymour$^{1,2}$, E.Breedt$^{1,3}$, P.Uttley$^{1}$, \and E.K\"{o}rding$^{4}$ and
V.Tudose$^{5,6,7}$\\ 
$^1$Department of Physics and Astronomy, University Southampton,
SO17~1BJ, UK \\
$^2$Mullard Space Science Laboratory, University College London, 
Holmbury St. Mary, Dorking, Surrey RH5 6NT, UK\\ 
$^3$Department of Physics, University of Warwick, Coventry,
CV4~7AL, UK \\ 
$^4$AIM/CEA Saclay and University Paris Diderot, l'Orme
des Merisers,F-91219 Gif-sur-Yvette CEDEX France\\ 
$^5$Netherlands Institute for Radio Astronomy, Postbus 2, 7990 AA 
Dwingeloo, The Netherlands\\
$^6$Astronomical Institute of the Romanian Academy, Cutitul de Argint 5, 
RO-040557 Bucharest, Romania\\
$^7$Research Center for Atomic Physics and Astrophysics, Atomistilor 
405, RO-077125 Bucharest, Romania\\}

\date{Accepted 2010 November 23.  Received 2010 November 23; in original form 2010 August 26}

\begin{document}

\maketitle

\begin{abstract}

We present intensive quasi-simultaneous X-ray and radio monitoring of
the narrow line Seyfert 1 galaxy NGC~4051, over a 16 month period in
2000-2001. The X-ray observations were made with the Proportional
Counter Array on the Rossi Timing X-ray Explorer (RXTE) and radio
observations were made at 8.4 and 4.8 GHz with the Very Large Array
(VLA).  In the X-ray band NGC~4051 behaves very much like the analogue
of a Galactic black hole binary (GBH) system in a `soft-state'. In
such systems, there has so far been no firm evidence for an active,
radio-emitting jet like those found in `hard state'
GBHs. VLBI observations of NGC~4051 show three well separated compact
components almost in a line. This structure resembles the core and
outer hot spots seen in powerful, jet-dominated, extragalactic radio
sources and, although no jet is visible in NGC~4051, suggests that a
weak jet may exist. However it has not previously been clear whether
the nucleus is currently active in the radio band and whether there is
any link between the radio and X-ray emission processes.

Radio monitoring of the core of NGC~4051 is complicated by the
presence of surrounding extended emission and by the changing array
configurations of the VLA. Only in the A configuration is the core
reasonably resolved. We have carefully removed the differing
contaminations of the core by extended emission in the various arrays.
The resulting lightcurve shows no sign of large amplitude variability
(i.e. factor 50\%) over the 16 month period and is consistent with being
constant. Within the 6 A configuration observations where we have
greatest sensitivity we see marginal evidence for radio core
variability of $\sim25$\% ($\sim0.12$mJy at 8.4GHz) on a 2-week
timescale, correlated with X-ray variations. These percentage
variations are similar to those of the Seyfert galaxy NGC~5548, which
is 10 times brighter.  Even if the radio variations in NGC~4051 are
real, the percentage variability is much less than in the X-ray band.
Within the B configuration observations, although sensitivity is
somewhat reduced, there is no sign of correlated X-ray/radio
variability.

NGC~4051 is one decade lower in radio luminosity than the
radio/X-ray/mass fundamental plane for jet-dominated
hard-state black holes, although the scatter around the plane is
of the same order. The lack of radio variability
commonly seen in hard state GBHs may be explained by orientation
effects.  Another possibility, consistent with the lack of radio
variability, is that the radio emission arises from the X-ray corona
although, in that case, the linear structure of the compact
radio components is hard to explain. A combination of corona and jet
may explain the observations.

\end{abstract}


\section{Introduction}

Galactic black hole X-ray binary systems (BHBs) are often classified
as being in one of a small number of `states'. The most common states
are the `hard' and the `soft' states. The hard state was originally
defined as the state where the 2-10 keV X-ray flux is low and the
X-ray spectrum is hard, and the `soft' state, where the 2-10 keV X-ray
flux is high and the X-ray spectrum is soft \citep{mcclintock06}. The
hard state is dominated by a power-law component, and the `soft' state
is characterized by a strong thermal blackbody component.  The main
physical parameter driving the states is the accretion rate with that
rate being higher in the soft than in the hard state \citep{Fenderjet,
Fenderjet3, Bellonistate}. Another major observational distinction
between the hard and the soft states is that compact, powerful
radio-emitting jets are usually present in the hard states
\citep{fender01,stirling01} but have not yet been detected in the soft
state. There have been detections of weak radio emission in soft state
systems \citep{brocksopp02,corbel04} but this emission is usually
thought to represent emission from relic plasma left over from a
previous hard-state jet and that no active jet is currently being
powered by the black hole \citep{Fenderjet2}.

NGC~4051 is a relatively nearby \citep[15.21 Mpc, ][]{russell02}
narrow line Seyfert 1 (NLS1) galaxy. It is one of the brightest and
most variable Active Galactic Nuclei (AGN) in the X-ray sky and has
been extensively observed by a number of X-ray observatories
\citep{mch04,Ponti06,Tera08,Elme2010}. One of the major results to
emerge from these studies is that the X-ray variability
characteristics of NGC~4051, as parameterised by the power spectral
density (PSD), are very similar to those of the Galactic black hole
X-ray binary (GBH) system Cyg X-1 when in the `soft' state
\citep{mch04}. A classification of NGC~4051 as a soft state system is
supported by the relatively high accretion rate ($\sim15\%$ \me)
\citep{Woo02}. If it were identical to a soft state GBH then an active
radio-emitting jet would not be expected. Although the X-ray
observations of other Seyfert galaxies are less extensive, the data
are consistent with Seyfert galaxies being soft state systems (e.g. also see \citet{mch05a}).

Radio variability has been detected previously in Seyfert galaxies but
observations are not extensive. \citet{Neff} detect variability of
$>20$\% in the Seyfert 2 galaxy Mkn 348 at 5GHz on $\sim$yearly
timescales and \citet{Wrobel} found variability at 8.4GHz in the
Seyfert 1 NGC~5548 of 33\% $\pm$ 5\% over 41 days and 52\% $\pm$ 5\%
over 4 years. \citet{Mundell} finds similar variations in most members
of a sample of mainly Seyfert 2 galaxies and \citet{Falcke2001} find
the same in most members of samples of radio-quiet and
radio-intermediate quasars. However in none of the above cases were
there any quasi-simultaneous X-ray observations. Recently \citet{Bell2010} have
studied the X-ray and radio variability of the liner NGC~7213 and find
a weakly significant correlation between those bands. Liners have much
lower accretion rates than Seyfert galaxies, being similar in
accretion rate to the hard state GBHs. Liners are also much more radio
loud than Seyfert galaxies, which is in agreement with the hard state
scenario, and so an X-ray/radio correlation is expected. In soft-state GBHs, 
as the radio emission has been scarcely detected, we 
do not know what the exact relationship to the X-ray emission might 
be. Correlated radio and X-ray observations of Seyfert galaxies may 
therefore provide us with important hints on the radio properties of 
the soft state systems

The soft state Seyfert galaxy NGC~4051 has been known for some time to
host a weak nuclear radio source of unknown origin
\citep[e.g.][]{Ulvestad84,kukula95,christopoulou97}.  Based on the
disappearance of the HeII 4686 line from the rms spectrum at low X-ray
flux levels, \citet{peterson2000} suggested that the inner part of
accretion disc may change from being optically thick to being
advective at low accretion rates, leading to the expectation of
increased radio emission \citep[e.g.][]{dimatteo99}. Therefore, over a
period of 16 months in 2000 and 2001, we carried out a joint radio
(with the Very Large Array, VLA) and X-ray (with the Rossi X-ray
Timing Explorer, RXTE) monitoring programme to search for an expected
anti-correlation between the radio and X-ray fluxes.

\begin{figure*}
\includegraphics[height=75mm]{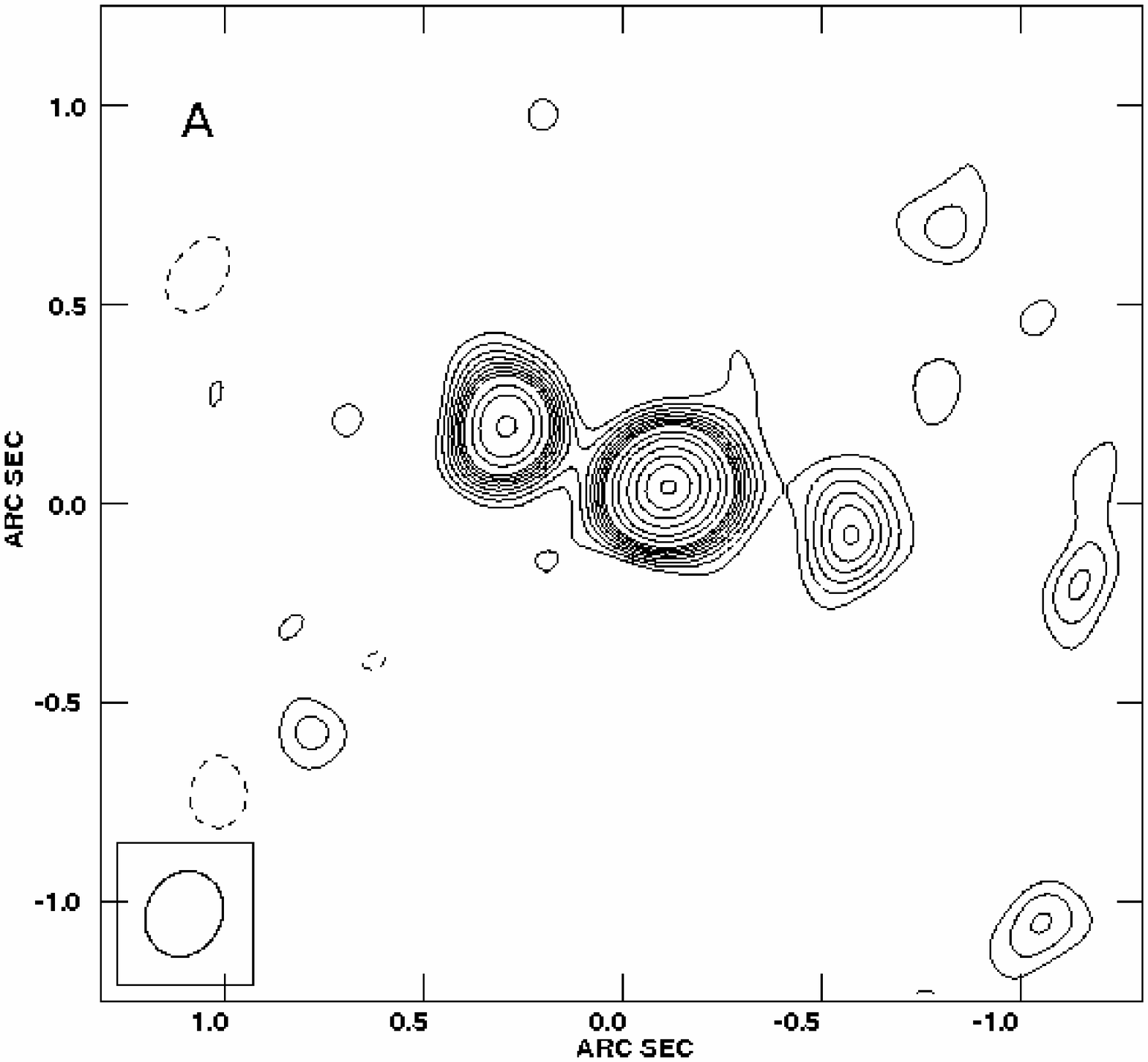} 
\includegraphics[height=75mm]{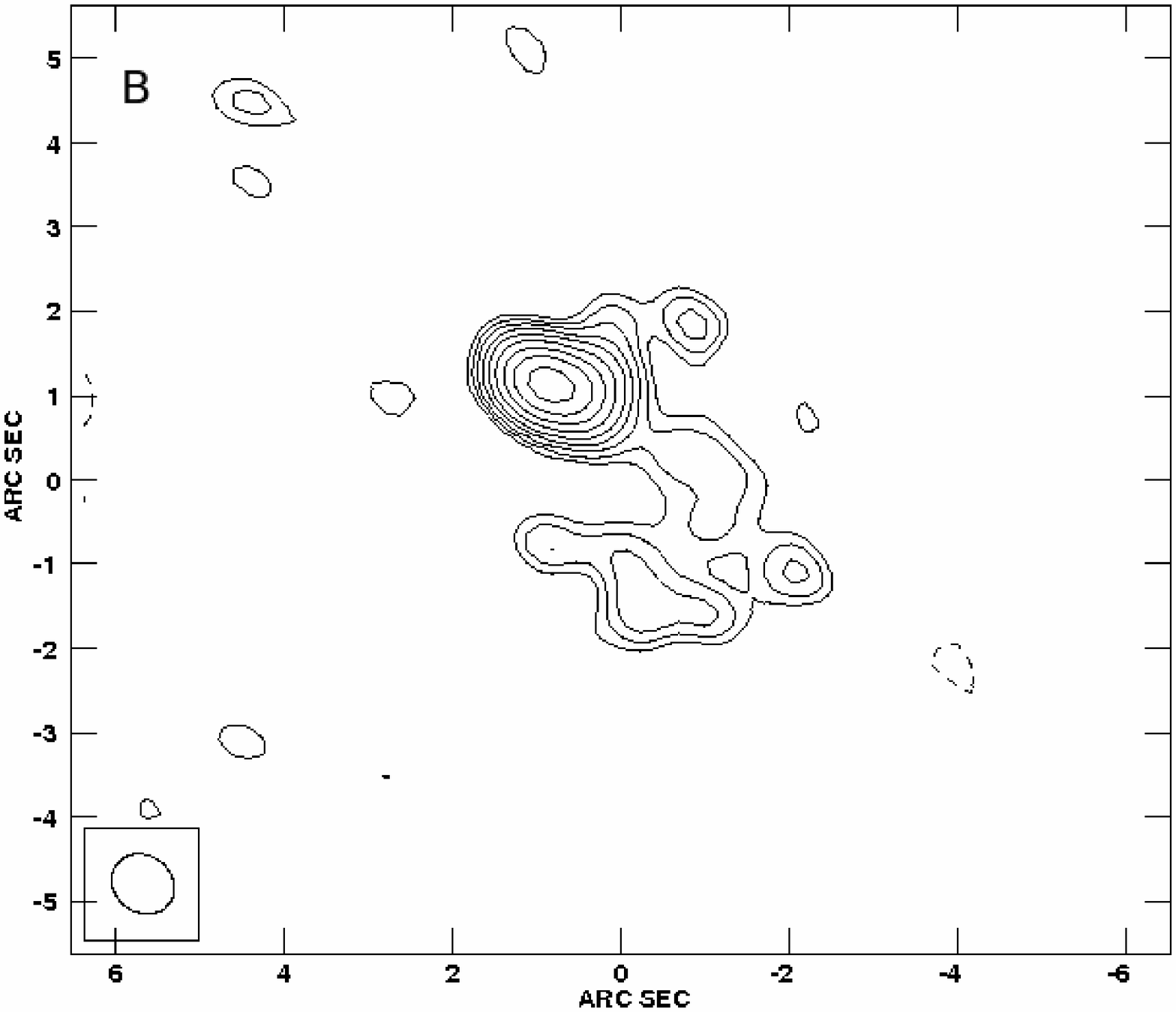} 
\includegraphics[height=75mm]{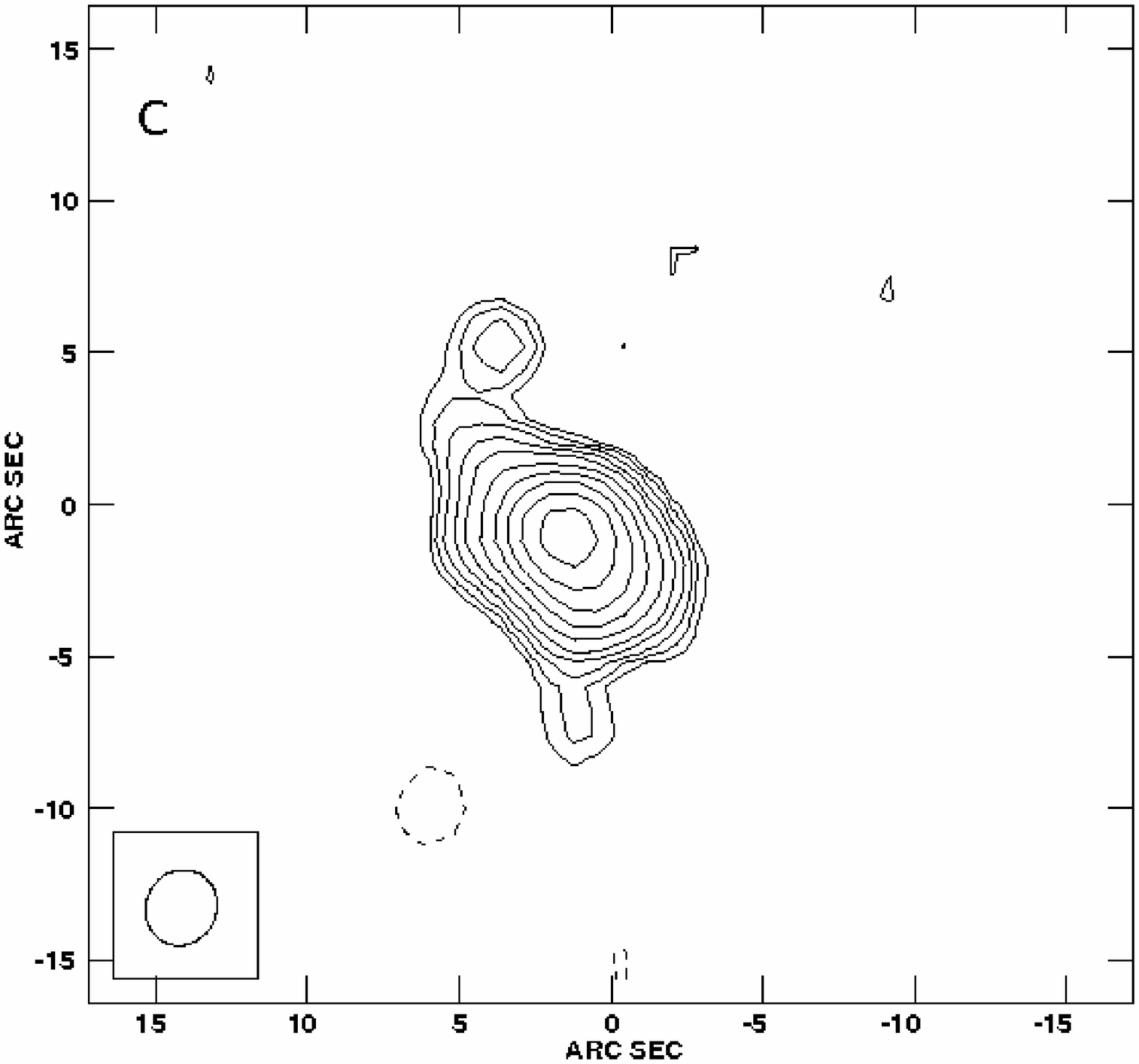} 
\includegraphics[height=75mm]{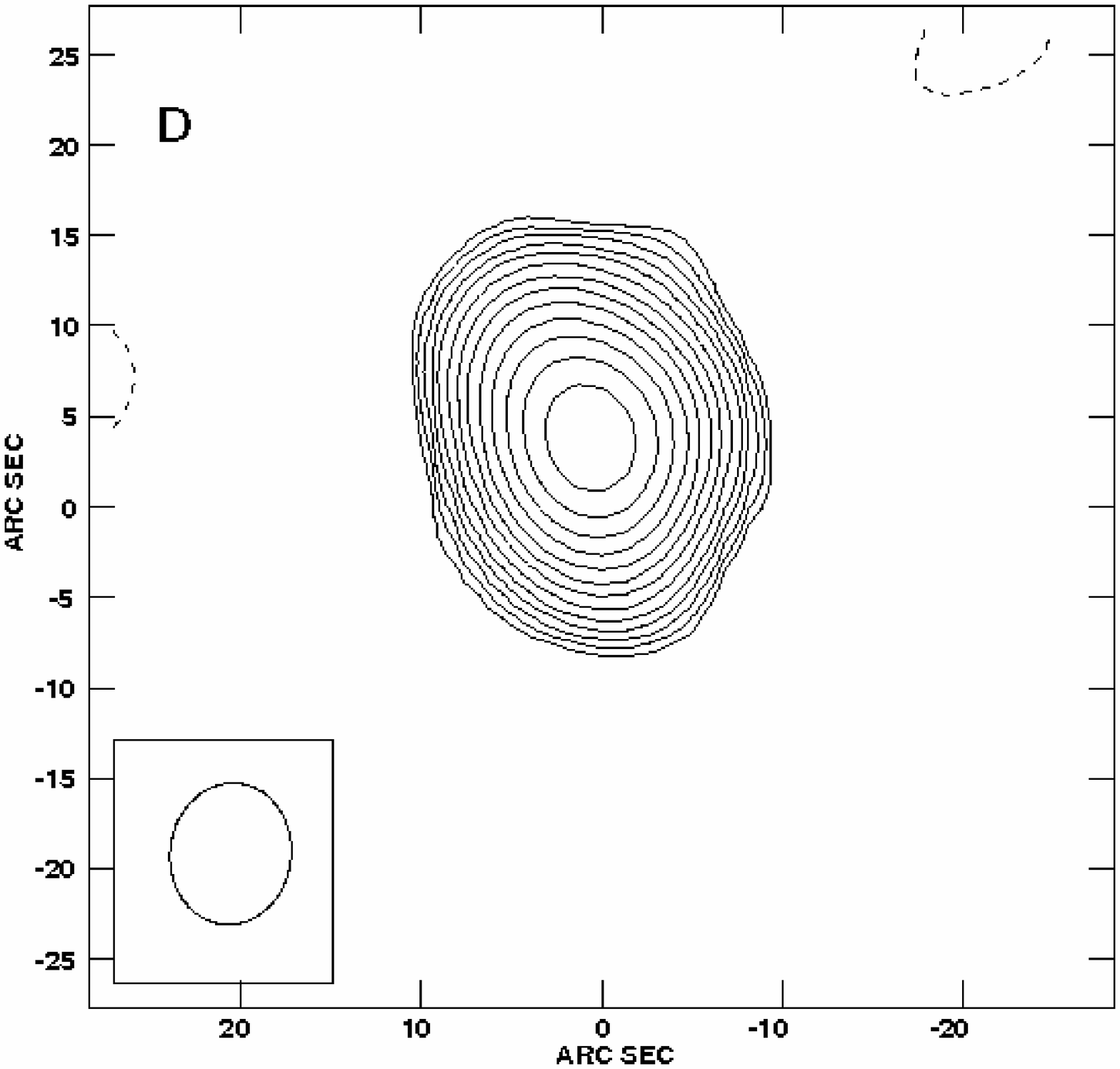} 
\caption{\protect\footnotesize Representative maps made from single 8.4GHz
  observations in each of the four main VLA array configurations. The
  configuration (A, B, C or D) is listed in the top left corner of
  each map. The maps are all centred on the nucleus of NGC~4051. The
  rms noise level in each A, B, C and D map is $16,2,2.2$ and
  $1.61\mu$Jy/beam and the major axis of the restoring beam is 0.22'',
  0.78'', 2.54'' and 7.83'' respectively . The contours are at
  $rms\times -2.8, 2.8, 4,5.6, 8, 11, 16, 23, 32, 45, 64, 90, 127,
  180, 254$.}
\label{fig1:4maps}
\end{figure*}

Preliminary analysis of the resultant data did not, however, reveal a
strong radio/X-ray anticorrelation. Instead the datasets appeared to
be positively correlated \citep{mch05b}. Correlated radio/X-ray
variability is commonly seen in hard state, jet-dominated, GBHs,
suggesting that the radio emission in NGC~4051 may also come from a jet. 
We therefore undertook radio Very Long
Baseline Interferometry (VLBI) observations both with the European
VLBI Network (EVN) in 2003 and the Global VLBI network in 2004. A
triple structure consisting of a compact core with a barely resolved
($\sim$ few milliarcsec) component on each side of the core, separated
by $\sim0.5$~arcsec and almost in a straight line, was detected
\citep{mch05b}. This structure is similar to the core and hot spots of
an FRII radio source and strongly suggests an underlying jet with the
hot spots being the places where the jet hits some surrounding medium.
Similar EVN observations were later performed by \citet{GiroVLBI}, revealing a
broadly similar structure.  A definitive analysis of the X-ray and
radio variability of the nucleus of NGC~4051 was, however, hindered by
the fact that the VLA array configuration changed approximately every
4 months and thus the size of the synthesised beam changed,
encompassing greater or lesser amounts of surrounding extended
emission (see Fig~\ref{fig1:4maps}). In this paper we carefully model
the radio emission of NGC~4051 on different spatial scales in order to
remove the contribution from extended emission to the flux from
the nucleus, which we then compare to the X-ray flux.

In Section 2 we present the radio observations and radio analysis. In
Section 3 we consider the radio variability as observed with the A and
B array configurations, in which sensitivity to core radio flux
variations is greatest, and compare it with quasi-simultaneous X-ray
observations. We find, at best, only a marginal relationship. In
Section 4 we discuss how to remove the contribution from extended
radio emission in the B, C and D array configurations. We find that,
although the X-ray emission varies by factors of 10 or more over the
16 month observation period, there is no believable long term
variation of the radio emission over the same period.
In Section 5
we show simulations designed to determine the accuracy with which the
extended contribution can be measured. In Section 6 we consider the
implications of our result for radio emission models for radio
quiet AGN. We include discussion of jet models which underly 
the so-called `fundamental plane' relating black hole mass to
radio and X-ray luminosity \citep{Falcke,merloni03} and we also discuss coronal
models \citep{laor08}.

\section{Radio Observations and data reduction}

NGC~4051 was observed by the VLA 29 times between 2000 16th June and
2001 3rd September with a typical interval between observations of 2
weeks. On each occasion the source was observed at both 8.4 and 4.8
GHz with a total time on source, in each band, of approximately 12
minutes. Six of our observations were made with the VLA in A
configuration, 2 in BnA, 6 in B, 1 in BnC, 6 in C, 3 in DnC and 5 in
D. We observed in 2IF mode with 50MHz bandwidth at both frequencies.
The phase calibrator was J1219+484 and the flux calibrator was
3C286. For more information on these observations see Table ~\ref{obs}.

The data were flagged and calibrated, using {\sc aips}, in the
standard manner.  Maps were made using {\sc imagr} with clean fields
set around two fainter neighbouring sources, one approximately 5.5
arcmin to the NE and another approximately 3.5 arcmin to the
south. After experimenting with different iterations of cleaning we
standardised on 10,000 iterations. Self calibration was not used on
the final images.  All early experimentation with self calibration
revealed the source to be too faint, and resulting images were not
improved by this method.

Our aim here is to search for a possible relationship between the
X-ray and radio fluxes. The X-rays vary very rapidly, and hence come
from a very small region, tens or hundreds of light seconds
across. If there is any relationship with the radio emission, we must
therefore expect the radio emission to come from a region which,
although possibly not as small as that of the X-ray emission region, is
unlikely to be light years across. At the distance of NGC~4051
($\sim15$Mpc), 1~arcsec$\sim70$pc and hence we are concerned here with
the core radio flux. However measurement of that flux accurately for a
source with any extended structure is far from trivial and, even for
an almost unresolved source, it is not easy, as we discuss further in
Section~\ref{aarray}.

The standard {\sc aips} task for measuring the peak and integral flux densities
is {\sc jmfit}. This task fits a gaussian model to the values within a
box drawn around the source and outputs the integral flux density (Jy)
values and peak flux density (Jy/beam) values. Although we are interested
in intrinsic core flux, measurements of both peak and intergral
flux densities have value for our study and so both are recorded and
listed later.

A box was placed around the nuclear core of NGC~4051 in each map using
the task {\sc cowindow}. The box size was different for each array
configuration and was chosen so it included all the nuclear flux. We
kept the box size constant when calculating flux density values within
the same array. {\sc jmfit} outputs errors depending on the gaussian
fitting to the nucleus of the source. These jmfit errors are then
combined in quadrature with the standard 5\% error on the flux density
value to calculate the error on each of the 29 flux readings.

\begin{table}
   \caption{Observational Details of VLA data at 4.8 and 8.4 GHz. For both frequencies the phase calibrator is J1219+484 and the flux calibrator
3C286. The phase calibrator was generally observed for $\approx$5 min and the flux calibrator for $\approx4$ min per run.}
  \begin{tabular}{@{}lcc}
     Date    & Configuration  & Time on Source\\
JD-2450000& & s \\
&&\\ 
     1701.583 & C  & 1130\\ 

     1722.514 & DnC & 1170\\ 
     1734.542 & DnC & 1130\\ 
     1741.505 & DnC & 1090\\ 

     1750.467 & D & 1210\\ 
     1756.535 & D & 1210\\ 
     1771.465 & D & 1100\\ 
     1803.335 & D & 1140\\ 
     1817.342 & D & 770\\

     1852.222 & A & 1150\\ 
     1865.185 & A & 1160\\ 
     1879.148 & A & 1140\\ 
     1893.131 & A & 1080\\ 
     1909.066 & A & 1090\\ 
     1929.926 & A & 1140\\ 

     1941.001 & BnA & 1000\\ 
     1956.956 & BnA & 1070\\  

     1985.710 & B & 1130\\ 
     2005.781 & B & 1120\\ 
     2023.773 & B & 1120\\ 
     2034.660 & B & 1150\\ 
     2046.649 & B & 1120\\ 
     2060.660 & B & 1210\\ 

     2078.464 & BnC & 1250\\ 

     2091.505 & C & 1140\\ 
     2101.560 & C & 1120\\ 
     2114.444 & C & 770\\
     2132.351 & C & 1130\\ 
     2156.272 & C & 1230\\ 
  \end{tabular}
\label{obs}
\end{table}

\section{Variability of the nucleus}

\subsection{During period of A configuration observations}
\label{aarray}

\subsubsection{8.4GHz Observations}

In Fig~\ref{fig1:4maps} we present images of NGC~4051 at 8.4GHz from
representative individual observations made in each of the four VLA
array configurations. We can see that, with the A configuration, the core is
separated from the neighbouring eastern and western components. Higher
resolution maps made from observations by MERLIN
\citep[e.g.][]{christopoulou97}) or the EVN \citep{GiroVLBI,mch05b}
show the same structure.
Thus the flux of the central component, as seen in the A configuration
observations, can be taken as reasonably representative of the nuclear
flux. In all other array configurations measurements of the nuclear
flux are contaminated by extended emission, which restricts
the conclusions which can be drawn from these data. We consider the problem of
contamination in more detail in Sections 3.2 and 3.3 but here consider
the specific concerns related to the A configuration observations.

The integral flux density is the desired measurement for slightly
extended, but otherwise isolated sources, such as one might find in
deep surveys. However the peak flux is generally the best measure of
the true core flux. If the source is compact and isolated but
there are significant phase errors in the calibration which move some
of the flux slightly to the side, then the integral flux density would
be a better measure. (For a very brief discussion of how one might
measure core fluxes see, for example, \citealt{mullin08} and papers
referred to therein.)  However if the core is surrounded by extended
emission then the integral flux density will be much more affected by
small changes in beamshape which may include different amounts of
extended emission. Our A configuration maps, although showing a core
which is close to being unresolved (eg Fig~\ref{fig1:4maps})
consistently indicate an extension, of approximately 0.1 arcsec, along
the line of the 3 main source components and so the problem of
extended emission exists even for the A configuration observations.

\begin{figure}
\rotatebox{90}{\includegraphics[height=84mm,width=84mm,angle=270]{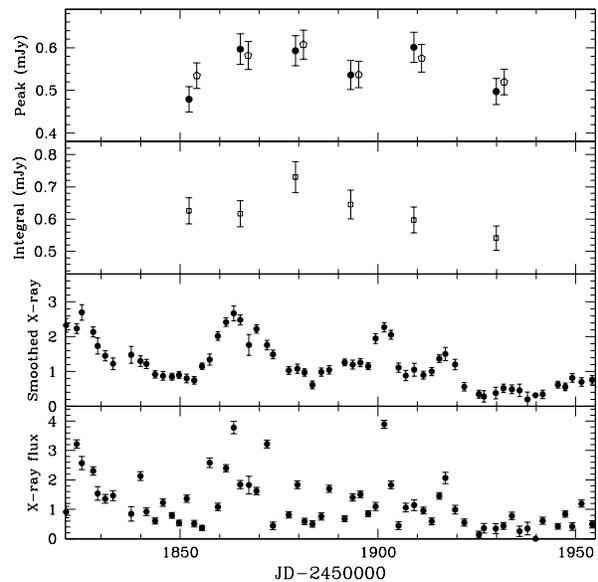}}
\caption{\protect\footnotesize Top panel: A configuration peak 8.4GHz core
flux densities with fixed beam (filled circles) and default/not fixed
beam (open pentagons). Note the non fixed beam points (open hexagons)
were moved forwards by 2 days to the make them easier to see. The flux
values were derived from {\sc jmfit} by fitting to maps made with both
fixed and default beams. See text for more details.  2nd panel: A
configuration integral intensity values for 8.4GHz core derived from
maps made with a non fixed restoring beam.  3rd panel: RXTE fluxes
slightly smoothed (running gaussian with a half-width 2 days) to remove high frequency
variability.  Bottom panel: 2-10 keV RXTE fluxes ($\times10^{-11}$ erg
cm$^{-2}$ s$^{-1}$). Observations occur approximately once every 2
days with typical duration 1ksec.}
\label{axrlc}
\end{figure}

In the top panel of Fig~\ref{axrlc} we show the 8.4GHz peak core flux
densities with fixed beam (filled circles) and default beams (open
hexagons) as derived using {\sc jmfit}. For the non fixed beam maps
the restoring beams were left free, to be determined only by the {\sc
UV} coverage and the use of robust 0 in {\sc imagr}. The size of the
fitted gaussian component was also left free. As the observations were
made within a relatively short period, at mostly the same hour angles,
the restoring beams, in practice, were very similar, although not
identical. The fitted gaussians were similar, although there was some
scatter in width.

In order to try and standardise further, we remade all the maps with a
fixed restoring beam with parameters at the mean of the unconstrained
restoring beam (i.e. for A configuration beamsize 0.24'' $\times$ 0.19''
and position angle of 137 degrees) and robustness 0.

The first point to note from Fig~\ref{axrlc} is that any variation of
the core 8.4GHz flux, if real, is very small, and certainly of much
lower amplitude than that seen in the X-ray observations (lower panels
of Fig~\ref{axrlc}). The quoted errors are the statistical error given
by {\sc jmfit} combined in quadrature with the 5\% error
usually assumed as the maximum likely uncertainty on the flux density
calibration. The calibration uncertainty therefore slightly dominates
the total resultant error which is typically twice the statistical
error, or slightly more. This total error is the correct error to use when
comparing the flux density from the same object in different
observations. To determine the relative flux densities of different
objects within the same observation, however, one would not include
the flux calibration uncertainty.

We also note that minor calibration errors, or sidelobes from distant
sources which have not been perfectly removed, can typically lead to
flux measurement errors which are larger by factors of two, or maybe
more, than the statistical error.  We can use observations of other
structures in the maps to make a crude estimate of the real errors on
the flux density of the central source. In our case components are
visible on either side of the core in the A configuration maps which
we might assume to be constant if they arise from extended components
related to NGC~4051. The (brighter) eastern component is definitely
more extended than the core, and the (fainter) western component is
probably the peak of a low surface brightness extended structure which
extends to the south and so flux measurements of both will be a little
more sensitive to the exact beamshape than will measurements of the
core. Nonetheless, the default beamshape does not change greatly but
variations in peak flux by twice the statistical error, and
occasionally by more, are seen between observations. We conclude that
the error which we use throughout this paper, i.e. the
statistical error from {\sc jmfit} combined in quadrature with a
5 percent amplitude calibration error , may be a slight underestimate of the true
total error but probably not by more than 30 percent.

We have also examined 8.4GHz A configuration observations made by other
observers. From an observation on 1991 June 24, \citet{kukula95} list
the core peak flux as 0.54 mJy and the integral flux as 0.60 mJy from
a map with rms noise level of 73 $\mu$Jy/beam. In an observation on
1991 September 01, which we analysed ourselves, we find a peak flux of
0.50 mJy with rms noise of 40 $\mu$Jy/beam (we do not quote the
integral flux density as the beam is more distorted than that of any
other A configuration maps which we consider here). These measurements are
very similar to those which we have measured in our own later
observations. 

We conclude that although we formally measure variations of $\sim25$
per cent on a timescale of 2 weeks between the first two observations
in the maps made with fixed restoring beam, the variation is barely
more than 0.1 mJy and, given the possible 30\% larger error, the
observations could be consistent with a constant source.

\subsubsection{8.4GHz to 4.8GHz spectral variations}

\begin{table}
   \caption{A configuration core peak flux density values at 8.4 GHz and
   4.8 GHz for maps made with fixed restoring beams. }
  \begin{tabular}{@{}lcccc}
     Radio    & Peak Flux Density  & Peak Flux Density \\
     Date     & at 8.4 GHz      & at 4.8 GHz  \\ 
    JD-2450000 &  $\times10^{-4}$ Jy &   $\times10^{-4}$ Jy\\
&&\\ 
     1852.222 & 4.79 $\pm$0.30 & 5.91  $\pm$0.36 \\ 
     1865.185 & 5.97 $\pm$0.36 & 6.36  $\pm$0.37 \\ 
     1879.148 & 5.93 $\pm$0.35 & 6.59  $\pm$0.38 \\ 
     1893.131 & 5.36 $\pm$0.34 & 6.41  $\pm$0.39 \\ 
     1909.066 & 6.01 $\pm$0.35 & 7.39  $\pm$0.43 \\ 
     1929.926 & 4.97 $\pm$0.31 & 5.98  $\pm$0.36 \\ 
  \end{tabular}
\label{tab1:A48}
\end{table}

\begin{figure}
\includegraphics[height=84mm,width=84mm]{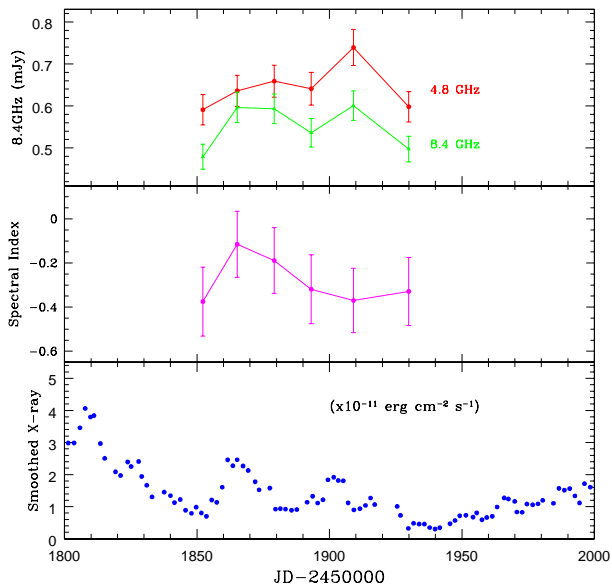}
\caption{\protect\footnotesize Upper panel: The core peak flux density (mJy) at 4.8 GHz
and 8.4 GHz during the A configuration derived from maps
made with identical restoring beams at each frequency.
Middle panel: The 2-point 8.4 to 4.8 GHz spectral index, $\alpha$
where $S(\nu) \propto \nu^{\alpha}$.
Bottom panel: Smoothed X-ray flux for the time period of the observations in A configuration. }
\label{spec}
\end{figure}

We carried out the same map making procedure on the
4.8GHz A configuration observations. In these cases we fixed the beam
size at 0.385 $\times$ 0.328 arcsec with position angle -15 degrees,
i.e. quite similar to the PA of the 8.4GHz observations. The resulting
peak flux densities are listed in Table~\ref{tab1:A48} and plotted in 
Fig~\ref{spec}. A broadly similar pattern of variations to that seen
at 8.4GHz is found and the amplitudes of variability are similar to
those seen in NGC~5548 by \citet{Wrobel}.

In Fig~\ref{spec} we also show the 8.4-4.8 GHz two-point spectral
index, $\alpha$, where $S(\nu) \propto \nu^{\alpha}$.  Given the
difficulty of measuring core fluxes we do not over-interpret these
results, however we do note that all the maps from which the fluxes
were derived were constructed in an identical way. With that proviso
we note that these observations reveal an average $\alpha= -0.2$,
similar to that found by \citet{christopoulou97}, with a slight
hardening at the beginning of the observations followed by a gradual
softening such as might be
explained by an injection of electrons with a relatively flat energy
distribution, e.g. perhaps as a result of a shock, followed by
radiative energy losses.

As this paper is concerned mainly with the radio variability and as
the problems of contamination of nuclear flux by extended emission are
even greater at 4.8 GHz than at 8.4 GHz, we do not consider the 4.8GHz
from configurations other than A configuration in this paper. However in the
A configuration the nucleus is still reasonably resolved at 4.8 GHz and so we
do present those data here. A full discussion of the spectral
morphology of NGC~4051 will be presented in a later paper.

\subsubsection{Radio/X-ray Correlation}

In Fig.~\ref{axrlc} we also show (bottom panel) the 2-10
keV X-ray fluxes derived from our monitoring programme with the Rossi
X-ray Timing Explorer (RXTE) using the Proportional Counter Array (PCA).
During the period of the A configuration
the RXTE observations typically occurred every 2 days, with a duration
of 1ksec. Description of the X-ray reduction techniques can be found
in other papers \cite[e.g][]{mch04}. It is well known that NGC~4051
varies very rapidly, and that rapid variability can be seen in
Fig.~\ref{axrlc}. Although there are slow long term
trends in the average flux of the source, observations with higher
time resolution by XMM-Newton \cite[e.g][]{mch04} reveal that
variations by a factor of 2 occur not uncommonly on timescales of an
hour. It is therefore relevant to consider how close in time the X-ray
and radio observations should be in order to be useful for any study
of correlated variability. 

The answer depends on what we think the physical mechanisms
responsible for the X-ray and radio emission are. For example, in
blazars where the variable emission in both the X-ray and lower
frequency bands is believed to come from shocks in a relativistic jet
oriented towards the observer, time separations of a few hours or less
are required \citep[e.g.][]{mch07a}. 
Although the VLBI radio morphology of three compact co-linear
components is most easily explained by the presence of an unseen jet,
the radio/X-ray ratio is much lower
than in typical blazars and a similar relativistic jet-like origin for
both emissions is unlikely. The currently most favoured, and
energetically simplest, paradigm for the X-ray emission is
Comptonisation of optical-UV disc photons by very hot thermal
($T>10^{9}$K) or non-thermal electrons in a corona above the accretion
disc \citep[e.g.][]{shapiro76}. One possibility for the radio emission
is that it is synchrotron emission from that same corona, although the
size of the radio emitting region at the frequencies considered here
would be typically 100 times larger than that of the X-ray emitting
region \citep[see][for an extensive discussion of this
model]{laor08}. In that case we would not expect to detect radio
variability on the rapid timescales seen in the X-ray band and it
would be more appropriate to look for correlations between the radio
emission and a longer-term averaged X-ray emission which may represent
longer term accretion rate fuelling of the overall emission regions.

In order to avoid pre-judging the issue, we have therefore
investigated the relationship between both the smoothed, and
unsmoothed, X-ray fluxes. (The X-ray lightcurve is only gently
smoothed with a running gaussian function of half-width 2 days.) 
The smoothed X-ray lightcurve is shown in the second panel
from the bottom in Fig.~\ref{axrlc}.

In case the time difference between the radio and X-ray observations
is important, we list the radio observation dates and the nearest RXTE
observation dates in Table~\ref{tab2:xray}. We list both the directly
observed, i.e. unsmoothed X-ray fluxes and the fluxes interpolated to
the time of the radio observations. We note that the greatest
separation is 0.59d but that three of the separations are within 3
hours. We note that, in fact, there is not a great deal of difference
between the nearest observed and interpolated X-ray fluxes.

\begin{figure}
\includegraphics[height=84mm,width=50mm,angle=270]{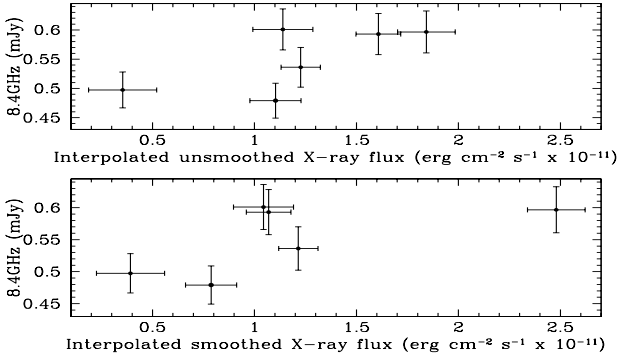}
\caption{\protect\footnotesize 8.4GHz A configuration peak flux densities derived from
  maps made with the same restoring beam plotted against the X-ray flux. 
  In the top panel we use
  the observed X-ray fluxes from the bottom panel of Fig.~\ref{axrlc},
  interpolated to the time of the radio observations. The exact time
  of the radio and nearest X-ray observations are given in Table
  2.  All radio observations have
  an X-ray observation within 0.6d and, in 3 cases, within 3 hours. In
  the bottom panel here we plot the X-ray fluxes interpolated from the
  slightly smoothed X-ray fluxes (third panel of Fig.~\ref{axrlc}).}
\label{asamebeam}
\end{figure}

\begin{figure}
\includegraphics[height=84mm,width=50mm,angle=270]{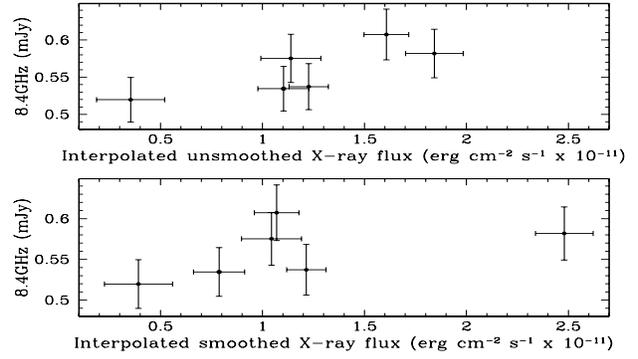}
\caption{\protect\footnotesize Peak 8.4GHz core flux densities derived from maps
  made with the default restoring beam in A configuration plotted against the X-ray flux. In the top panel we use the observed unsmoothed X-ray fluxes, and in the bottom panel we use the smoothed X-ray fluxes.}
\label{ajmpk}
\end{figure}

\begin{figure}
\includegraphics[height=84mm,width=50mm,angle=270]{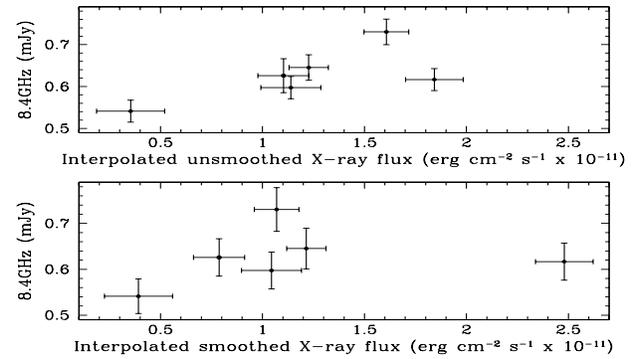}
\caption{\protect\footnotesize Integral Intensity radio flux densities at 8.4GHz core flux derived from maps
  made with the default restoring beam in A configuration plotted against the observed X-ray fluxes (top panel) and the smoothed X-ray fluxes (bottom panel) .}
\label{ajmint}
\end{figure}

\begin{figure}
\includegraphics[height=84mm,width=50mm,angle=270]{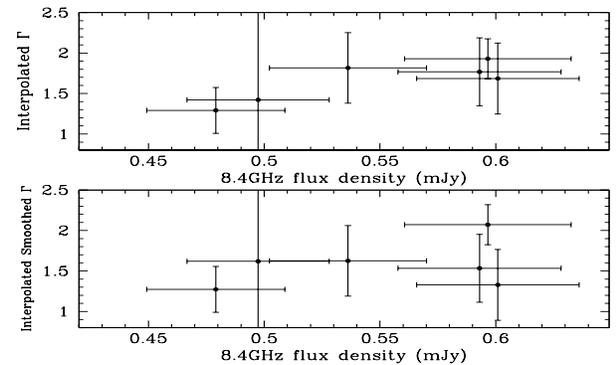}
\caption{\protect\footnotesize Peak 8.4GHz flux densities from A configuration maps with the same
  restoring beam plotted against RXTE 2-10 keV photon number spectral index,
  $\Gamma$. In the top panel we interpolate from a slightly smoothed lightcurve of $\Gamma$ and in the bottom panel we simply interpolate $\Gamma$ between
  the nearest observed values. }
\label{xspec}
\end{figure}

In Fig.~\ref{asamebeam} we plot the 8.4GHz radio flux
from the maps made with the same restoring beam against both the
interpolated unsmoothed and interpolated smoothed X-ray fluxes. There
is a weak correlation between the radio and X-ray fluxes although we
repeat our caution that, with a slightly larger error, the radio
fluxes could be consistent with being constant. Smoothing the X-ray
flux does not improve the appearance of the relationship. In 
Figs.~\ref{ajmpk} and \ref{ajmint} we plot the 
same relationship for the peak, and integral, 8.4GHz flux densities
derived from the maps made with the default restoring beams. The same
general patterns are observed. We are able to measure simple powerlaw
X-ray spectral indices for each of the RXTE observations and, in 
Fig.~\ref{xspec} we plot the interpolated photon number indices, from
both smoothed and unsmoothed lightcurves as before, against radio flux
density for the 8.4GHz maps made with the same restoring beam. Again a
similar pattern is seen although we caution that there is usually a
good relationship between X-ray flux and spectral index for Seyfert
galaxies such as NGC~4051 \citep{lamer03_4051} so these plots are not
completely independent of the previous flux-flux plots.

Hardening of radio spectral with increasing flux followed by
softening with decreasing flux can be simply explained in the context
of jet models by the injection of particles with a hard spectrum
followed by subsequent radiative losses, and so might explain our
observations of radio variability. However if the X-ray emission also
arose predominantly in the jet we would expect a similar spectral
behaviour, which is not what we see here. We refrain from speculating
too much given the quality of the data but the correlated variability
seen here, if real, may simply reflect changes in fuelling rate to
both emission regions.

\begin{table}
   \caption{8.4~GHz observation dates with the nearest RXTE X-ray
   observation dates. X-ray fluxes are given both as observed and
   interpolated to the time of the radio observation.
Dates are JD-2450000. X-ray fluxes are in units
   of $10^{-11}$ erg cm$^{-2}$ s$^{-1}$.}
  \begin{tabular}{@{}lccc}
     Radio    & Interpolated   & X-ray    & Observed   \\ 
     Date     & X-ray flux     & Date     & X-ray flux    \\
              &                &          &                \\
     1852.222 & 1.10 $\pm$0.13 & 1851.631 & 1.37 $\pm$0.13 \\ 
              &                & 1853.551 & 0.51 $\pm$0.11 \\ 
     1865.185 & 1.84 $\pm$0.14 & 1865.178 & 1.84 $\pm$0.14 \\ 
     1879.148 & 1.61 $\pm$0.11 & 1879.632 & 1.83 $\pm$0.13 \\ 
     1893.131 & 1.23 $\pm$0.10 & 1893.637 & 1.41 $\pm$0.13 \\ 
     1909.066 & 1.14 $\pm$0.15 & 1909.184 & 1.14 $\pm$0.19 \\ 
     1929.926 & 0.35 $\pm$0.17 & 1929.794 & 0.35 $\pm$0.17 \\
  \end{tabular}
\label{tab2:xray}
\end{table}

\subsection{During period of B configuration observations}
\label{barray}

\subsubsection{8.4GHz observations}

We see, from Fig.~\ref{fig1:4maps}, that in the B configuration
observations the core is not resolved from the two close components on
either side. Measurement of the core flux therefore becomes more
difficult and flux density measurement errors will be larger than for
the A configuration observations. Nonetheless it is still possible to
measure the peak, and integrated, flux densities.

\begin{figure}
\includegraphics[width=89mm]{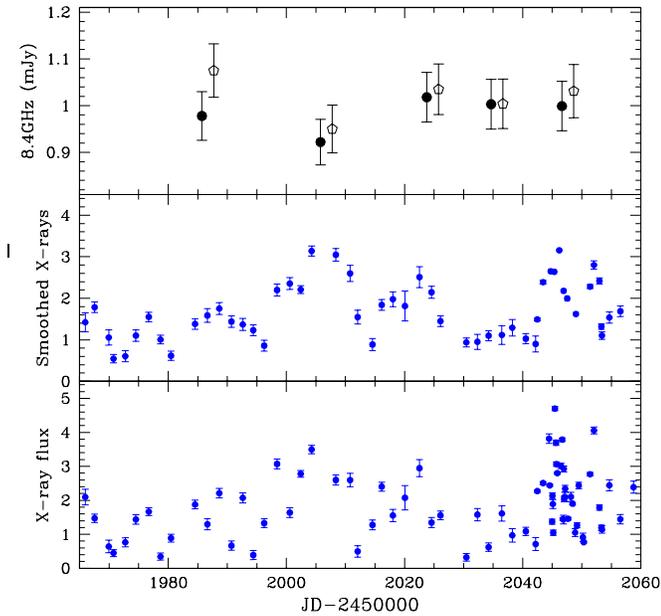}
\caption{\protect\footnotesize In the top panel we plot the peak core flux
  densities at 8.4GHz derived from B configuration maps made with the
  default/unfixed restoring beam (open pentagons) and fixed beam (filled
  circles). Note, the unfixed beam points have been increased by 2
  days, to stop overlap and make them easier to see. We plot these
  against the unsmoothed and smoothed X-ray flux values or the period
  of the B configuration observations in the second and third panel
  respectively.}
\label{bxrlc}
\end{figure}

In Fig~\ref{bxrlc} we show the lightcurves during the period of the B
configuration observations.  Here the radio flux densities used are
both the peak core flux densities measured from maps made with the
default beamshape and those with a fixed beam . As with the A
configuration observations, the values derived from maps made with the
same/fixed restoring beam (i.e. 0.76 $\times$ 0.64 arcsec and position
angle of -25.13 degrees) are quite similar to those made with the
default beam.  The X-ray observations are as
described for the A configuration observations. We again see, at most,
only marginal radio variability with the 8.4 GHz flux density
consistent with being constant.

\subsubsection{Radio/X-ray correlation}

\begin{figure}
\includegraphics[height=84mm,width=60mm,angle=270]{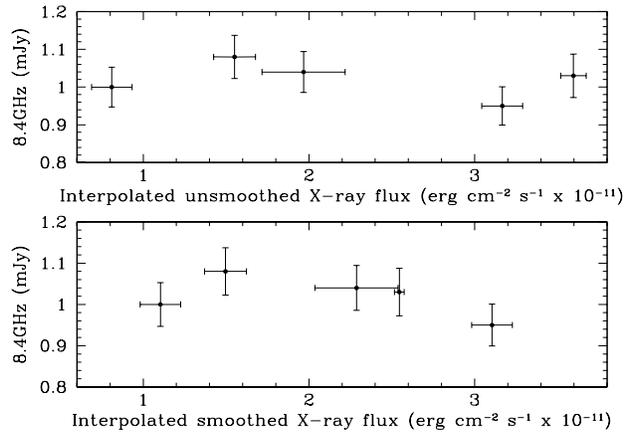}
\caption{\protect\footnotesize Peak radio flux densities of the core at 8.4GHz  derived from maps made with the default restoring beam in B configuration plotted against the observed X-ray fluxes (top panel) and the smoothed X-ray fluxes (bottom panel).}
\label{bjmpk}
\end{figure}

The relationship between the radio and X-ray fluxes in the B configuration
 is shown in Fig.~\ref{bjmpk}, again displaying both the smoothed and unsmoothed
X-ray fluxes. The hint of a positive X-ray/radio correlation seen in
the A configuration observations is not seen here.
With the possible exception of the highest X-ray
flux point, there is no sign of any variation of the radio flux
despite considerable variation in the X-ray flux. Unfortunately that
highest X-ray flux point is derived from the largest extrapolation of fluxes
as the RXTE observation closest to that time had zero useful
exposure. Thus we have had to extrapolate between two observations
which were 4d apart.

\subsection{All configurations}

\begin{figure}
\includegraphics[width=90mm,height=84mm]{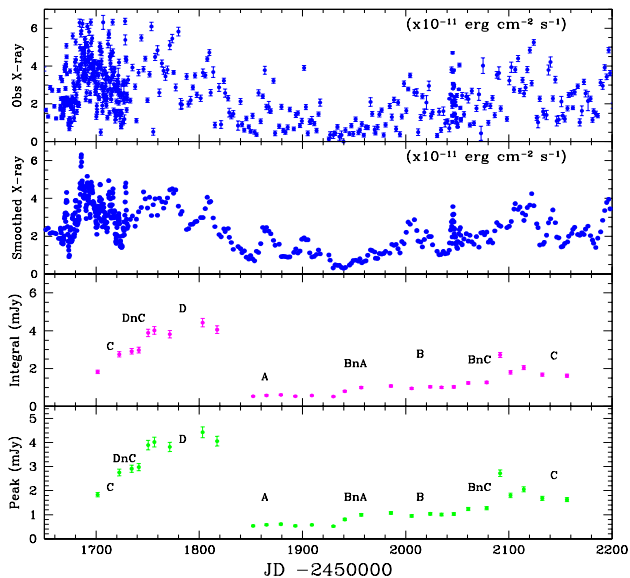} 
\caption{\protect\footnotesize Top panel - directly observed X-ray fluxes. Second panel - X-ray flux smoothed with a 4 point boxcar. Third panel -integral flux densities at 8.4GHz peak labelled with VLA array configuration. Fourth (Bottom) panel - peak flux densities at 8.4GHz peak labelled with VLA array configuration. The apparent variability in the radio flux densities is largely due to the changing array configuration of the VLA every $\sim~100$ days.}
\label{rawxray}
\end{figure}

Although the errors increase, it is possible to measure peak
and integrated flux densities also for the C and D array configurations.  In
Fig~\ref{rawxray} we plot the resultant 8.4 GHz fluxes densities for
all configurations together with the observed, and slightly smoothed,
X-ray fluxes. Within the individual array configurations there is no
evidence of strong variability although, over months timescales, it
appears visually as if there is a strong correlation between the radio
and X-ray fluxes.

Unfortunately, by chance, the period of the lowest X-ray flux occurs
during the A configuration observations and the period of largest
X-ray flux occurs during the D configuration observations.  However
note that here we are plotting the observed radio flux densities and
have not yet removed contributions to the core flux density from
extended emission.  As the nuclear radio flux will be most
contaminated by extended emission during the D configuration
observations, it is probable that at least part of the apparent
correlation between the radio and X-ray fluxes is actually a
consequence of the changing array configuration. In order to properly
compare the radio and X-ray variability we must therefore remove the
extended radio flux, as we describe below.

\section{Comparing radio fluxes between arrays}

The procedure for removing the contribution to the core flux density
from extended emission is carried out in stages, offsetting by one
array configuration change at a time. For example, in offsetting from
B configuration to A configuration we first determine, from real
observations with B configuration, what the typical beam shape and
size is in B configuration. We then take the A configuration
observations and taper the {\sc uv} datasets, using {\sc uvrange}
within {\sc imagr}, so that the beam sizes are the same as they would
be in B configuration. 
We refer to the resulting map as a pseudo B configuration map.
We then measure both the
peak and integral flux densities in the pseudo B array map and
determine the differences, or offsets, between values measured from
the original, untapered, A configuration map.  We repeat the process
for all of the A configuration datasets and determine the average offsets to
the B configuration, and an rms error thereon, for the peak and integral
flux densities.

To offset the real observed B configuration observations to an A
configuration observation we remove the offset derived above from each
of the observed B configuration flux densities and we combine the rms error on
the offset with the observational error on the B configuration flux densities
in quadrature.

\begin{figure}
\includegraphics[height=84mm,width=84mm]{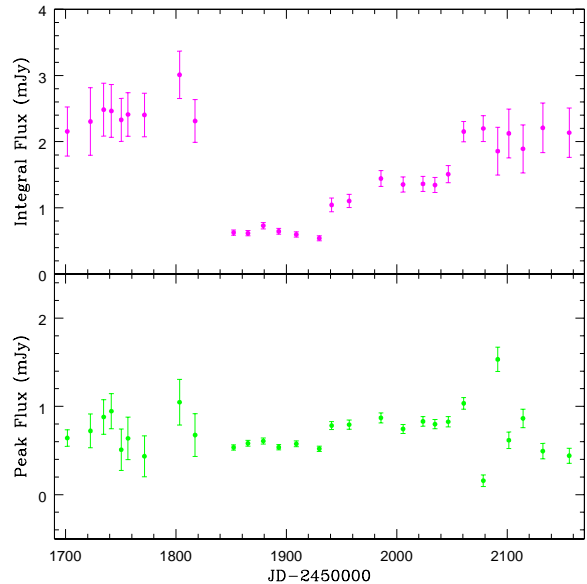}
\caption{ Integral flux density (top panel) and peak flux density (bottom panel) at 8.4GHz of the core. Flux densities for all observations are offset to the A array configuration. The peak flux density is consistent with being constant. }
\label{offset}
\end{figure}

We perform this offset method on the other configurations to get average offset values for D to C, and C to B.
Combining these offsets with the offset from B to A allows us to offset all data to an A array configuration. 
For example, the average peak flux density offset in
going from A to B configuration is 0.32$\pm 0.07$mJy and in going from C to D
configuration it is 2.31$\pm 0.21$mJy. In offsetting over more than one
array configuration change we combine the offsets linearly and the
errors in quadrature. Thus these average offset values were used to
offset all flux density values to those of an equivalent A configuration
observation. The resultant offset peak and integral flux density
lightcurves are shown in Fig.~\ref{offset}. In the intermediate
configurations such as BnC there are few epochs of observation, and
the configuration is slightly different in all of them. It is
therefore very difficult to estimate the offset. These observations
give rise to the more deviant peak flux densities. However even
including these points the peak flux densities are consistent with
being constant as long as we assume the more realistic larger error.  For
the integral flux densities, although the variations are now
considerably reducted, there is still some residual variability, which
still follows the X-ray variability. However we
note that the changes still mainly follow the changing array
configurations. Within all configurations apart from the A
configuration, there is no evidence for variability so one must
suspect an error in the offsetting procedure for integral flux
densities. In the next section we simulate the offsetting
method for integral flux densities to determine its accuracy.

\section{Simulating the Offset Method}

The first step in simulating the offset method was to make a model of
the source.  Although it is preferable that the model resembles the
real source, it is not essential that it is identical to the real
source as the aim of the simulations is only to test the offsetting
method and could, in principle, be carried out on any extended source.

\begin{figure}
\includegraphics[height=84mm]{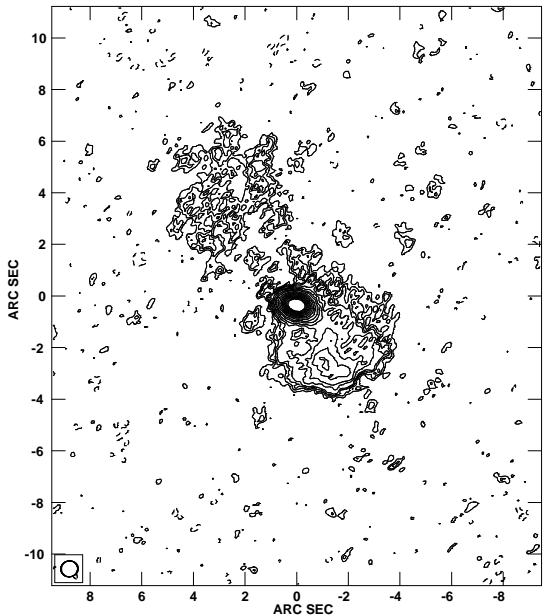}
\caption{\protect\footnotesize Image made by combining the UV data from all single arrays combined at 8.4GHz and restored with a typical
B configuration beam (0.65'' $\times$ 0.6''). The rms noise level
is~$1.103\times10^{-05}$ Jy/beam and contours are at the following multiples of
the noise level -2, 2, 3, 4, 5, 6, 7, 8, 9, 12, 15, 20, 25, 30, 35, 40, 45, 50, 55, 60. }
\label{fig7:model}
\end{figure}

\begin{figure}
\includegraphics[height=84mm]{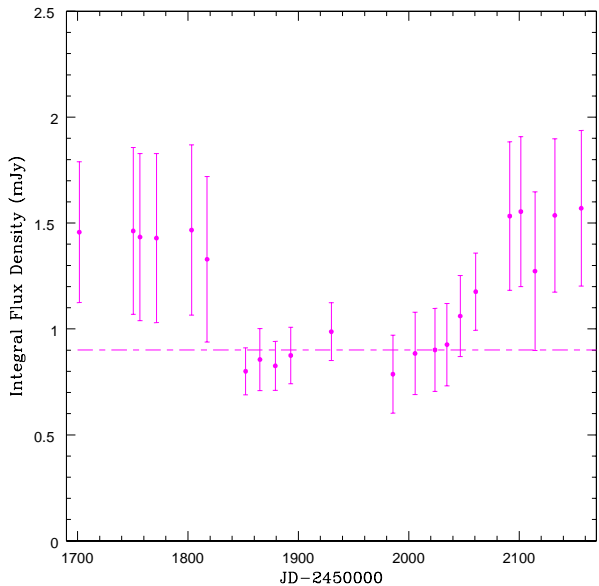} 
\caption{\protect\footnotesize Integral flux densities for the model NGC~4051 data, offset
to A array. Here the mean A configuration model value is 0.9 mJy (shown by the dotted line).
In offsetting from array to array, offsetting errors are added in quadrature to the {\sc jmfit} error and the 5\% error on the flux. We note that the
B configuration model data, when offset to A are at the expected A array flux values,  but the mean C and D configuration data differ, on average, from the
expected offset value by $\sim0.5$ mJy.}
\label{fig8:simulated_lc}
\end{figure}

To make the model we combined all the {\sc uv} data sets from the A,
B, C \& D configurations into one {\sc uv} data set which we imaged
(Fig.~\ref{fig7:model}).  The 28 clean components from this combined
image which best represented the structure of the AGN were then
selected.  Using {\sc uvsub} we added the model contribution from
these components to the residual {\sc uv} data sets produced by {\sc
imagr} during our original imaging of the 22 observations in each of
the 4 main array configurations. The resultant data sets represent how
our actual observational setups would have seen our model of NGC~4051.
We then repeated the offsetting procedure, which we described above
for the real data, on the simulated data, to produce a simulated
integrated flux density offset to the A configuration.  This procedure should
result in a constant flux density level equal to the model A configuration
values. The actual simulated offset lightcurve is shown in
Fig.~\ref{fig8:simulated_lc}.  The dotted horizontal line in
Fig.~\ref{fig8:simulated_lc} represents the expected A array integral
flux density value.  We see that the B configuration data are at the same
level as the A configuration data but the C and D configuration data are too high by
about 0.5-0.6~mJy which therefore represents an approximate systematic
error in our process of offsetting the integral C and D configuration flux
densities. If this systematic error is subtracted from the original
offset integral flux densities (Fig.~\ref{offset}), then most of the
apparent long term variation is removed from the integral flux
densities. It is unfortunate that the array configuration changes
follow the approximate changes in long term X-ray flux, and hence that
the very small residual variation in the integral flux densities
appears to follow the X-ray flux. However given the particular
difficulties of measuring the integral flux densities, the fact that
the changes closely follow the array changes and the fact
that long term variability is not seen in the peak flux densities, we
conclude that the apparent small residual long term variability in the
integral flux densities does not represent real variation of the core flux.

\section{Discussion}

\subsection{Radio Variability}
We have monitored the radio and X-ray variability of the probable soft
state Seyfert galaxy NGC~4051 over a period of 16 months in 2000 and
2001 with observations approximately every 2 weeks. We have carefully
subtracted the contributions to the core flux density from the
differing amounts of extended flux density which will be detected in
the different restoring beams in the different VLA configurations.
Thus we have produced lightcurves equal to those which would have been
produced by observations all with the A array.
Over that period the core peak flux density which in this context is the best measurement of true core radio luminosity, is consistent with being
constant. Also, the average peak radio flux density during our observations
is, within the errors, the same as during two previous observations in
June and September of 1991. 

Within the A configuration, which is the only configuration within
which the core is resolved from close neighbouring components to the
east and west, there is a hint of correlated X-ray and radio
variability, with radio (and X-ray) spectral variations approximately
following the flux variations.  However although the X-ray flux varies
by factors of a few, the maximum radio variations are, at most, 25
percent between observations separated by $\sim2$ weeks. However the
amplitude of variability is, at most, 0.12mJy. Whilst the percentage
changes, timescales of change, and spectral hardening during a rise
are entirely consistent with the observations of radio variability of
another similar Seyfert galaxy, NGC~5548 \citep{Wrobel}. NGC~5548 is
approximately ten times brighter than NGC~4051 and so there is no
doubt about the credibility of its radio variability. Unfortunately
for NGC~4051 the credibility of the variability depends strongly on
the error associated with each measurement.  Taking the larger of the
errors which could be associated with the radio measurements
(discussed earlier) the radio lightcurve would be consistent with
being constant. Thus overall we conclude that there is no strong
evidence for any large amplitude variability in the core radio flux of
NGC~4051 at 8.4GHz over the course of a year but we cannot absolutely
rule out very low amplitude ($\sim0.1$mJy) variability correlated
with the X-ray variability.

\begin{figure}
\includegraphics[width=80mm,height=60mm]{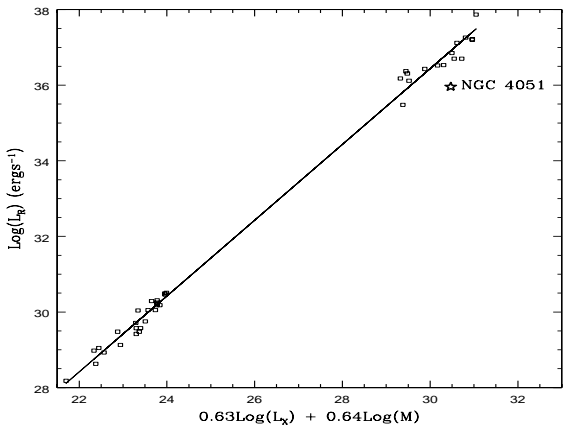}
\caption{\protect\footnotesize The radio/X-ray/black hole mass `fundamental plane' for
  hard state accreting black holes (i.e. hard state X-ray binary systems
  and liners) using the data from the \citet{KFC06} sample. NGC~4051
  is the large black star, and is labelled.}
\label{kfcfund}
\end{figure}

\begin{figure}
\includegraphics[width=80mm,height=60mm]{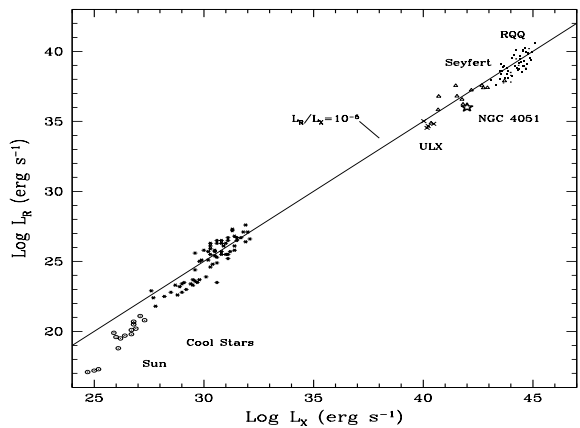}
\caption{\protect\footnotesize Our mean values of X-ray and peak radio luminosity for
  NGC~4051 (large 5-pointed star) superposed on the lower panel from
  Fig.2 of \citet{laor08}. 
The solid line is the G$\ddot{u}$del-Benz relation, i.e.
  $L_{R}/L_{X}=10^{-5}$, for coronally active stars \citep{gudel93}. 
}
\label{laor}
\end{figure}

\subsection{Implications for emission models}

In Fig.~\ref{kfcfund} we plot our values of the radio luminosity and
the average X-ray luminosity on the so-called `fundamental plane' of
black hole accretion activity \citep{merloni03,Falcke}. Here we plot a
version of the plane using only `hard state' black holes from the
sample of \cite{KFC06}. We note that NGC~4051, which is almost
certainly a `soft state' object, lies about a decade below (in the sense of
being more radio quiet) the best-fit relationship defined by the hard
state objects, although it is just about within the scatter of the
relationship.  However hard state objects typically show much more
radio variability than NGC~4051 and are typically described by $L_{R}
\propto L_{X}^{\beta}$ where $\beta \sim 0.7$ for variations in
individual objects \citep{corbel03,gallo03,gallo06}. For NGC~4051
$\beta=0$, or possibly $\beta=0.1$ during the A configuration observations.

\citet{Bell2010} find a significant correlation
between X-ray and radio emission in the LINER NGC~7213 with factors of
$\sim 2$ or more variability in both bands. Given its very
low accretion rate, NGC~7213 is thought to be a hard state system
so correlated radio/X-ray variability is expected. However although
\citet{Wrobel} has demonstrated radio variability in the Seyfert galaxy
NGC~5548, which has an accretion rate not far below that of NGC~4051
and so is probably also a soft state system, there have been no
previous studies of correlated X-ray/radio variability in
similar Seyfert galaxies. The X-ray/radio
relationship has, however, been investigated for samples of
objects. \citet{panessa07} find $\beta=0.97$ at 4.8GHz and
$\beta=0.98-1.25$ at 15GHz for a sample of Seyfert galaxies. For a
sample of ROSAT-selected AGN, and using 1.4GHz radio observations,
\citet{brinkmann00} find $\beta=0.48 \pm 0.049$ for radio quiet
objects, similar to Seyferts, (and $\beta=1.012 \pm 0.083$ for radio
loud objects). Thus although there is a wide range in the reported
values of $\beta$, there is good evidence for radio/X-ray correlation
within samples of Seyferts and radio quiet AGN. However within the
only well studied 'soft state' Seyfert so far, i.e. NGC~4051, there is
no evidence for a value of $\beta$ much above 0.1. 

One possibility is that even in the VLA A configuration 8.4GHz maps, the core
still contains substantial contribution from emission which is
extended but on scales below that detectable with the VLA. We note
that with a beam of $\sim8$mas, \citet{GiroVLBI} find an unresolved
source with 5GHz flux density of $0.20 \pm 0.02$mJy, which is
substantially less than our average 4.8GHz measurement of
$\sim0.6$mJy. Our tentative observations of $\sim 0.1$mJy variability
would then provide a value of $\beta$ consistent with the values
derived from observations of samples of similar sources, although NGC~4051
would then be even more radio quiet compared to the hard state objects
which define the fundamental plane. 

The VLBI observations show structure which is strongly indicative of a
collimated jet \citep{GiroVLBI, mch05b}. (In a future paper we will
present our higher resolution global VLBI observations which confirm
this structure.) Thus as first sight the lack of radio variability is
surprising. However if the jet is in the plane of the sky, rather than
pointing at the observer as in blazars, radio variability would be
slower. \citet{falcke01} speculate similarly regarding the lack of
strong radio variability in radio-loud quasars although they do not
reach any firm conclusions.

Another alternative is that the radio emission in Seyfert galaxies
does not come from a jet, as in hard state sources, but arises in the
outer parts of the X-ray emitting corona \citep{laor08}. In
Fig.~\ref{laor} we plot our current data onto Fig.2 from
\citet{laor08} and see that NGC~4051 agrees reasonably with the G$\rm
\ddot{u}$del-Benz \citeyear{gudel93} relationship for coronally active
stars, i.e. $L_{R}/L_{X}=10^{-5}$.  In the coronal model, rapid large
amplitude radio variability is not expected at frequencies where the
source is optically thick. This situation is approximately the case
during our A configuration observations which indicate a spectral
index of $\alpha \sim -0.2$.  Such variability is, indeed, not
seen. Similar considerations would, of course, also apply to some
extent to a side-on jet and the highly collimated co-linear structure
of the compact VLBI components is a significant problem for the
coronal model.  As jets are generally agreed to arise from the corona,
the truth may lie part way between the pure coronal and pure jet
models. In this galaxy where an outflow of large solid angle is seen
in [OIII] observations \citep{christopoulou97}, the compact radio
components separated from the nucleus may arise from a central, higher
velocity region within the overall outflow.  The core radio emission
may arise from a combination of a jet and a corona. If there is any
long term correlated variations between the radio and X-ray bands,
they may simply reflect long term changes in accretion rate, upon
which both bands depend, rather than a direct link between the
emission processes.

\section{Conclusions}

Despite factors of 10 X-ray variability, we see no clear evidence
for variability of the core of the soft state narrow line Seyfert 1
galaxy NGC~4051 at 8.4GHz over a 16 month period of monitoring at
2-weekly intervals, with the possible exception of very low amplitude
($\sim0.12$mJy) variations during the A configuration observations
where the core is best resolved from surrounding structures. The
latter tentative variations correlate weakly with the much larger
amplitude X-ray variations. Our resultant radio and mean X-ray
luminosity make NGC~4051 about a decade radio quieter than the hard
state objects which define the `fundamental plane' for hard state
accreting black holes, although the scatter about the plane is almost
of the same order. Given the collimated VLBI structure which hints at the
presence of an unseen jet, the lack of radio variability is, at first sight,
surprising, although a side-on jet, observed at a frequency where it
is optically thick, would not vary as rapidly as a face-on blazar
jet. A coronal model agrees well with the radio/X-ray flux ratio, and
the lack of radio variability, but the collimated radio structure is
then hard to explain. A mixture of jet and coronal emission may
explain the observations but further, more sensitive, radio
observations, with high angular resolution and a fixed beamshape, are
required to confirm whether the radio emission from NGC~4051 does
indeed vary.\\

{\bf ACKNOWLEDGMENTS}\\

We thank Ari Laor, Julia Riley, Tom Muxlow, Martin Bell and an
anonymous referee for useful discussions and comments. We also thank
Ari Laor for providing the Supermongo code for producing Fig.2 of
\cite{laor08}. SJ acknowledges support from an STFC research
studentship and IMcH acknowledges support under STFC grant
ST/G003084/1

\bibliographystyle{mn2e} 


\end{document}